\documentclass[twocolumn,amsmath,amssymb]{revtex4}
\usepackage{amsmath}
\usepackage{extarrows}
\usepackage[ruled]{algorithm2e}
\usepackage{graphicx}
\usepackage{dcolumn}
\usepackage{bm}
\usepackage[all]{xy}
\usepackage{indentfirst}
\usepackage{multirow}
\usepackage{mathrsfs}
\usepackage{bbm}
\usepackage{euscript}
\usepackage{amssymb}
\usepackage{extarrows}
\usepackage{bbm}
\usepackage[ruled]{algorithm2e}
\usepackage{graphicx}
\usepackage{epstopdf}
\usepackage{dcolumn}
\usepackage{bm}
\usepackage[all]{xy}
\usepackage{indentfirst}
\usepackage{amsmath}
\usepackage{multirow}
\usepackage{mathrsfs}
\usepackage{euscript}
\usepackage{amssymb}
\usepackage{appendix}
\usepackage{color,xcolor}
\usepackage{txfonts}

\newtheorem{theorem}{Theorem}
\newtheorem{definition}{Definition}
\newtheorem{proposition}{Proposition}
\newtheorem{lemma}{Lemma}

\begin{document}

\title{A new entanglement measure based dual entropy}

\author{Xue Yang, Yan-Han Yang, Li-Ming Zhao, Ming-Xing Luo}

\affiliation{The School of Information Science and Technology, Southwest Jiaotong University, Chengdu 610031, China
}

\begin{abstract}
Quantum entropy is an important measure for describing the uncertainty of a quantum state, more uncertainty in subsystems implies stronger quantum entanglement between subsystems. Our goal in this work is to quantify bipartite entanglement using both von Neumann entropy and its complementary dual. We first propose a type of dual entropy from Shannon entropy. We define $S^{t}$-entropy entanglement based on von Neumann entropy and its complementary dual. This implies an analytic formula for two-qubit systems. We show that the monogamy properties of the $S^{t}$-entropy entanglement and the entanglement of formation are inequivalent for high-dimensional systems. We finally prove a new type of entanglement polygon inequality in terms of $S^{t}$-entropy entanglement for quantum entangled networks. These results show new features of multipartite entanglement in quantum information processing.
\end{abstract}

\maketitle

\section{Introduction}

Entanglement as one of the most fascinating phenomena in quantum mechanics distinguishes the quantum world and the classical one. The ubiquity of entangled quantum states as a resource is well-known, and their effectiveness for applications often hinges on the degree of entanglement in the quantum state. Entanglement is usually quantified by the entropy of entanglement that arises from a subsystem while the information about the remaining system is ignored \cite{3H2009}. Entanglement entropy has been used to probe the various properties of many-body systems \cite{ZhangY2011,Isakov2011,Abanin2012,Islam2015,Barghathi2018,Zhao2022} or nonlocality of quantum networks \cite{TALR}.

It is well-known that quantum entropy is a valuable quantity for describing the uncertainty of a quantum state.  For a pure state $|\psi\rangle_{AB}$, it can be quantified via the von Neumann entanglement entropies \cite{Nielsen}:
\begin{eqnarray}
S(\varrho_A)=-{\rm Tr}\varrho_A\log_2\varrho_A,
\label{S0}
\end{eqnarray}
where $\varrho_A={\rm Tr}_B(\rho_{AB})$ is the reduced density operator of subsystem $A$ obtained by tracing out the subsystem $B$ and $\rho_{AB}=|\psi\rangle_{AB}\langle\psi|$ is the density matrix of the whole system. This way for quantifying bipartite entanglement follows by the idea that more uncertainty in subsystems implies stronger quantum entanglement between subsystems. A myriad of measures of entanglement entropy have so far been proposed such as the concurrence \cite{Hill1997}, the entanglement of formation (EOF)\cite{Bennett19963824}, R\'{e}nyi-$\alpha$ entropy entanglement \cite{HHH1996,Gour2007,Kim2010R}, Tsallis-$q$ entropy entanglement \cite{LV1998,Kim2010T}, and Unified-$(q,s)$ entropy entanglement \cite{KimBarry2011}. Howbeit, among these entanglement measures, there exists a common defect that bipartite entanglement is defined as the entropy of only one subsystem. This follows a natural problem is what more can be learned from a given state beyond von Neumann entropy.

One of the most fundamental issues concerned with the entanglement entropy measure is the limited shareability of bipartite entanglement for multipartite entangled systems. Followed by the original spirit of the Coffman-Kundu-Wootters (CKW) inequality \cite{V.Coffman}, the entanglement distribution for a three-qubit system is firstly displayed analytically as the following form:
\begin{eqnarray}
 E_{A|BC}\geq  E_{A|B}+ E_{A|C},
\label{eqn0}
\end{eqnarray}
where $E_{A|BC}$ is an entanglement measure of a composite quantum system containing systems $A, B, C$  under the bipartition $A$ and $BC$, $E_{A|B}$ and $E_{A|C}$ are entanglements of bipartite systems. This relation has been re-called as the monogamy of entanglement \cite{V.Coffman,Terhal2004}. Later, Osborne and Verstraete extend the monogamy inequality with the squared concurrence for any $n$-qubit systems \cite{T.J.Osborne}. Similar monogamy inequalities for multi-qubit states hold with the squared EOF \cite{Oliveira2014,Bai3,Bai2014}, the squared R\'{e}nyi-$\alpha$ entropy \cite{R2015}, the squared Tsallis-$q$ entropy \cite{Luo2016},  and the squared Unified-$(r, s)$ entropy \cite{Khan2019}. Interestingly, a set of tight $\alpha$-th powers monogamy relations have been investigated for multi-qubit systems \cite{Zhu2014,Luo2015,Luo2016}. Interestingly, the inequality (\ref{eqn0}) even the equality may be hold for entangled quantum network \cite{Luo2022} while it is invalid for specific high-dimensional systems \cite{Luo2021}. And only one known the squashed entanglement is monogamous for arbitrary dimensional systems \cite{Christandl2004}. The traditional monogamy inequality (\ref{eqn0}) provides a lower bound for ``one-to-group'' entanglement, such ``one-to-group'' entanglements are also named quantum marginal entanglements \cite{Walter2013}. There is another kind of entanglement distribution relation giving an upper bound for quantum marginal entanglements as the following polygon inequality \cite{Qian2018}:
\begin{eqnarray}
 E_{A|BC}\leq  E_{B|AC}+ E_{C|AB}
\label{eqn00}
\end{eqnarray}
for any tripartite entangled pure states. It can be regarded as resource sharing rules for distributed quantum applications \cite{Qian2018}. Notably, high-dimensional entanglement opens intriguing perspectives in quantum information science \cite{Friis2019} or quantum communications \cite{Cerf2002,Sheridan2010,Mafu2013,Mirhosseini2015,Islam2017,Cozzolino2019}.  Moreover, it may provide a platform for fundamental research in technological advances \cite{Erhard2020}. Hence, a natural problem is to consider the monogamy relations and polygon inequalities for higher dimensional systems.

The outline of the rest is as follows. In Sec.II, we define the total entropy of the von Neumann entropy and its  complementary dual. We propose some properties of the new entropy such as the nonnegativity, concavity, symmetry, additivity, and boundedness. In Sec.III, we establish a type of bipartite entanglement measure named as $S^{t}$-entropy entanglement. This gives an analytic formula for two-qubit entangled systems. In Sec.IV, we investigate the monogamy of qubit systems and higher-dimensional systems in terms of $S^{t}$-entropy entanglement. We show the monogamy properties of $S^{t}$-entropy entanglement and entanglement of formation (EOF) are inequivalent in higher-dimensional cases. We present an entanglement polygon inequality for arbitrary $n$-qudit quantum network states in  Sec.V while the last section concludes the paper.

\section{The total entropy of von Neumann entropy and its complementary dual}

Consider a discrete random variable $X$ on sample set $\{x_1, x_2,\cdots,x_n\}$. Its probability distribution is given by $p_i=Pr(X=x_i)$. This implies a self-information about each event $x_i$ as $I(X=x_i)=-\log_2p_i$. The average information of the random variable $X$ is then known as Shannon entropy \cite{Shannon} defined by
\begin{eqnarray}
H(X)=-\sum^n_{i=1}p_i\log_2p_i.
\label{entropyc0}
\end{eqnarray}
This gives the uncertainty which has before the statistical experiment. Now, for a given event $x_i$ one may define a new binomial random variable $X_i\sim\{p_i,1-p_i\}$, that is, by regarding all the events except for $x_i$ as a complementary dual event of $x_i$. In this case, we get its self-information as $-\log_2(1-p_i)$. This follows a new binary entropy as
\begin{eqnarray}
H(X_i)=-p_i\log_2p_i-(1-p_i)\log_2(1-p_i)
\label{Eqn0}
\end{eqnarray}
for $i=1, \cdots, n$. Moreover, all the complementary dual information are framed into an entropy function as
\begin{eqnarray}
\overline{H}(X)=-\sum^n_{i=1}(1-p_i)\log_2(1-p_i).
\label{entropyc00}
\end{eqnarray}
This is firstly introduced as ``extropy'', a  complementary dual of Shannon entropy  \cite{Lad2015}.

In classical information theory, the intuitive link between the uncertainty and information is that the greater the uncertainty of the given distribution, the more information gain from learning the outcome of the experiment. This means the Shannon entropy only presents the information all we know concerning which event will occur while ignores the information of its complement. However, the extropy (\ref{entropyc00}) can make up for this weakness. Thus, we redefine the entropy of the random variable $X$ as
\begin{eqnarray}
H^{t}(X)=\sum^n_{i=1}H(X_i)=H(X)+\overline{H}(X).
\label{entropyc1}
\end{eqnarray}

\begin{figure}[htb]
\begin{center}
\resizebox{240pt}{170pt}{\includegraphics{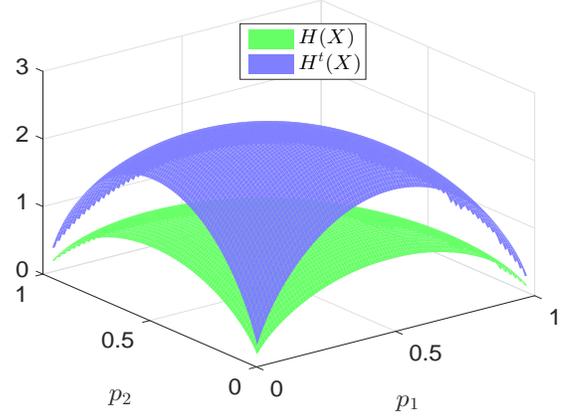}}
\end{center}
\caption{\small (Color online). A comparison of the Shannon entropy $H(X)$ and total entropy $H^{t}(X)$ with respect to the probability distribution $\{p_i\}$.}
\label{HSg}
\end{figure}

\emph{Example 1}. Consider a random variable $X$ with probability distribution $p_i=Pr(X=x_i)$ associated with $X=x_i, i=1, 2, 3$, where $0\leq p_i\leq1$ and $\sum^3_{i=1}p_i=1$. The Shannon entropy is given by
\begin{eqnarray}
H(X)=-\sum^3_{i=1}p_i\log_2p_i,
\end{eqnarray}
while the total entropy is shown as
\begin{eqnarray}
H^{t}(X)=-\sum^3_{i=1}p_i\log_2p_i-(1-p_i)\log_2(1-p_i).
\end{eqnarray}
As its illustrated in Fig.\ref{HSg}, the total entropy $H^{t}(X)$ shows more uncertainty beyond Shannon entropy.

In quantum scenarios, inspired by the von Neumann entropy we can define the quantum total entropy for a given quantum state.

\begin{definition}
The total entropy of a quantum state $\rho$ on $d$-dimensional Hilbert space ${\cal H}$ is defined by
\begin{eqnarray}
S^{t}(\rho)=-{\rm{Tr}}[\rho\log_2\rho+(\mathbbm{1}-\rho)\log_2(\mathbbm{1}-\rho)],
\label{entropy1}
\end{eqnarray}
where $\mathbbm{1}$ is the identity matrix.

\label{Def1}
\end{definition}

For a given state $\rho$, the spectrum decomposition is given by $\rho=\sum_i\lambda_i|\psi_i\rangle\langle \psi_i|$, where $\lambda_i$ are eigenvalues associated the orthogonal eigenvectors $|\psi_i\rangle$. With this decomposition the $S^{t}$-entropy is rewritten into
\begin{eqnarray}
S^{t}(\rho)=-\sum^d_{i=1}(\lambda_i\log_2\lambda_i+(1-\lambda_i)\log_2(1-\lambda_i)).
\label{entropy11}
\end{eqnarray}
This states that the $S^{t}$-entropy of a density matrix $\rho$ is equal to the classical  $S^{t}$-entropy of the probability distribution $\vec{\lambda}=(\lambda_1,\lambda_2,\cdots, \lambda_d)$ from the its eigenvalues.

\begin{figure}[htb]
\begin{center}
\resizebox{240pt}{180pt}{\includegraphics{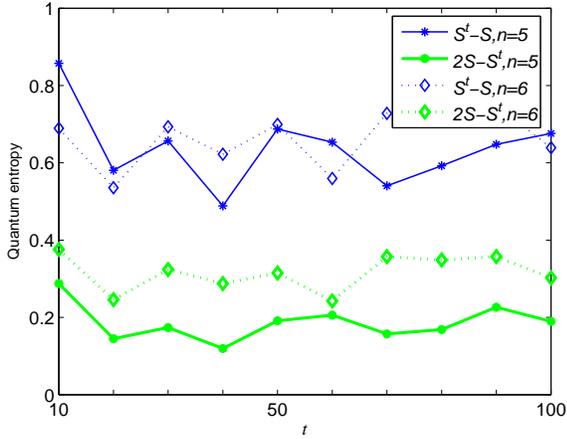}}
\end{center}
\caption{\small (Color online)  A comparison of von Neumann entropy and the total $S^t$ entropy in terms of evolution time $t$.}
\label{compare}
\end{figure}

\textit{Example 2}. Consider a $5$-qubit Heisenberg model with a random magnetic field in the $z$-direction. Its Hamiltonian is given by
\begin{eqnarray}
H_5&=&0.5\vec{\sigma}_1\cdot{}\vec{\sigma}_4
+0.4\vec{\sigma}_2\cdot{}\vec{\sigma}_3+
0.3\vec{\sigma}_3\cdot{}\vec{\sigma}_4
\nonumber
\\
&&
-0.5\vec{\sigma}_4\cdot{}\vec{\sigma}_5+\sum_{i=1}^5h_j\sigma_j^z,
 \end{eqnarray}
where we suppose that the pairs of $(1,3),(2,3),(3,4),(4,5)$ are correlated each other, herein, $\vec{\sigma}_j=(\sigma^x_j,\sigma^y_j, \sigma^z_j)$ represents a vector of Pauli matrices on qubit $j$, $h_j\in [-1,1]$ denotes the strength of the disorder. For a 6-qubit system, the Hamiltonian is given by
\begin{eqnarray}
H_6&=&0.4\vec{\sigma}_1\cdot{}\vec{\sigma}_3
+0.5\vec{\sigma}_2\cdot{}\vec{\sigma}_5-
0.3\vec{\sigma}_3\cdot{}\vec{\sigma}_4
\nonumber
\\
&&+0.2\vec{\sigma}_3\cdot{}\vec{\sigma}_6
+0.6\vec{\sigma}_5\cdot{}\vec{\sigma}_6
+\sum_{i=1}^6h_j\sigma_j^z,
\end{eqnarray}
where the pairs of $(1,3),(2,5),(3,4),(3,6),(5,6)$ have correlation with each other. Consider the initial state $|+\rangle^{\otimes 5}, |+\rangle^{\otimes 6}$ respectively, and the evolution time $100$, where $|+\rangle=\frac{1}{\sqrt{2}}(|0\rangle+|1\rangle)$. We can get the von Neumann entropy and the total $S^t$ entropy as shown in Fig.\ref{compare}.
It follows that the von Neumann entropy is strictly smaller than the total $S^t$ entropy while the total $S^t$ entropy is no more than twice of von Neumann entropy, that is, $S\leq S^t\leq 2S$.

Intuitively, the von Neumann entropy of a bipartite state $|\Phi\rangle_{AB}$ on Hilbert space $\mathcal{H}_A\otimes \mathcal{H}_B$ can be viewed as the uncertainty of an observer $A$ who can only access to the subsystem $A$. While similar to classical scenarios, the present $S^{t}$-entropy provides a method for describing its complementary dual information. In applications, one may suppose a mixed state represent an ensemble of pure states, that is, $\{p_i,|\psi_i\rangle\}$. In this case, the density operator is defined by $\rho=\sum_ip_i|\psi_i\langle\psi_i|$. Now, for each measurement outcome $i$  under the measurement operator $M_i=|\psi_i\rangle\langle\psi_i|$, the quantum statistical probability is $p_i={\rm Tr}(M_i\rho)$. We can regard all the measurements except for $M_i$ as the complementary dual measurement of $M_i$. This implies a positive-operator-value measurement (POVM) $\{M_i,I-M_i\}$, which follows the average information (\ref{Eqn0}). Hence, the total entropy $S^{t}$ gives an extended extropy information obtained from quantum statistics beyond the von Neumann entropy.

For a quantum system described by a density operator $\rho$, a general entropy function \cite{Canosa2002} may be defined as
\begin{eqnarray}
S_f(\rho)={\rm Tr}f(\rho)=\sum_if(\lambda_i),
\label{general0}
\end{eqnarray}
where $f$ is a smooth concave function satisfying $f(0)=f(1)=0$, $\lambda_i$ are the eigenvalues of $\rho$. This includes the von Neumann entropy with
\begin{eqnarray}
f(\lambda)=-\lambda\ln \lambda.
\end{eqnarray}
While the Tsallis entropy \cite{Tsallis1988} may be defined according to
\begin{eqnarray}
f(\lambda)=\frac{(\lambda-\lambda^q)}{q-1},
\end{eqnarray}
where $q>0$ and $q\neq1$. Instead, the present $S^{t}$-entropy (\ref{entropy11}) may be  recovered according to
\begin{eqnarray}
f(\lambda)=-\lambda\ln \lambda-(1-\lambda)\ln (1-\lambda).
\end{eqnarray}
This allows us construct one-parameter total entropy based on the Tsallis entropy and its complementary dual (see Appendix B).

Next, we show that the present $S^{t}$-entropy satisfies the following properties.

\begin{lemma} Let $\rho_{AB}$ be any density operator of a composite system $AB$, and $\varrho_{A(B)}$ be the reduced density operator of the subsystem $A(B)$. The total entropy $S^{t}$ satisfies the following properties.
\begin{itemize}
\item[(i)] \textbf{Non-negativity}: The total entropy satisfies $0\leq S^{t}(\rho)\leq d\log_2d-(d-1)\log_2(d-1)$.

\item[(ii)] \textbf{Concavity}: For any convex combination $\rho=\sum^n_{i=1}p_i\rho_i$, the total entropy satisfies
\begin{eqnarray}
\sum^n_{i=1}p_iS^{t}(\rho_i)\leq S^{t}(\rho).
\end{eqnarray}

\item[(iii)] \textbf{Symmetry}: For any pure state $\rho_{AB}$ the total entropy satisfies $ S^{t}(\varrho_A)=S^{t}(\varrho_B)$.

\item[(iv)] \textbf{Invariance}: The total entropy is invariant under any unitary transformation, that is, $S^{t}(\rho)=S^{t}(U\rho U^\dag)$ for any $U\in \mathbb{SU}({\cal H}_A\otimes{\cal H}_B)$.

\item[(v)]  \textbf{Subadditivity}: For any product state $\rho=\varrho_{A}\otimes\varrho_{B}$, the total entropy satisfies
\begin{eqnarray}
 S^{t}(\rho)< S^{t}(\varrho_A)+S^{t}(\varrho_B)
 \label{subadditivity}
\end{eqnarray}
and
\begin{eqnarray}
S^{t}(\rho)\geq \max\{S^{t}(\varrho_A),S^{t}(\varrho_B)\}.
\label{trace0}
\end{eqnarray}

\item[(vi)] \textbf{Boundedness}: For any convex combination $\rho=\sum^n_{i=1}p_i\rho_i$, the total entropy satisfies
\begin{eqnarray}
S^{t}(\rho)\leq H^{t}(X),
\label{bound0}
\end{eqnarray}
where $X$ is the classical random variable with distribution $\{p_1, \cdots, p_n\}$.

\end{itemize}

\label{Lemma1}
\end{lemma}

The proof of Lemma 1 is shown in Appendix A.

\section{The $S^{t}$-entropy entanglement}

Let ${\cal S}^{AB}={\cal S}(\mathcal{H}_{A}\otimes\mathcal{H}_B)$ be the set of density matrices acting on Hilbert space $\mathcal{H}_{A}\otimes\mathcal{H}_B$. A function $E$: $S^{AB} \to {\mathbbm R}_+$ is called a measure of entanglement if it satisfies the following conditions \cite{entanglement2001,entanglement2009}:
\begin{itemize}
\item[(E1)]$E(\rho)=0$ if and only if $\rho$ is a separable state;
\item[(E2)]Invariance under local unitary transformations: $E(\rho)=E((U_{A}\otimes U_{B})\rho (U^{\dag}_{A} \otimes U^{\dag}_{B}))$ for any local unitary transformations $U_{A}\in \mathbb{SU}(\mathcal{H}_{A}), U_B\in \mathbb{SU}(\mathcal{H}_B)$;
\item[(E3)] Non-increasing on average under local operations and classical communication (LOCC) operations that is, $\sum_jp_j E(\sigma_j)\leq  E(\rho)$, where $\Lambda^{LOCC}$ is an arbitrary LOCC operation defined by    $\Lambda^{LOCC}(\rho)=\sum_jp_j\sigma_j$.

\item[(E4)] Convexity: $E(\sum_ip_i\rho_i)\leq \sum_ip_i E(\rho_i)$ for any $\rho=\sum_ip_i\rho_i$, where $\{p_i\}$ is a probability distribution, and $\rho_i$'s are density matrices.
\end{itemize}

In what follows, we propose new bipartite entanglement measure named the $S^{t}$-entropy entanglement from the total quantum entropy.

\subsection{The $S^{t}$-entropy entanglement}

For a pure state $|\Phi\rangle_{AB}$ on $d\times d$ dimensional Hilbert space ${\cal H}_A\otimes {\cal H}_{B}$, the $S^{t}$-entropy entanglement is defined as
\begin{eqnarray}
E_t(|\Phi\rangle_{AB})=\frac{1}{r}S^{t}(\varrho_A),
\label{Eg1}
\end{eqnarray}
where $r=d\log_2d-(d-1)\log_2(d-1)$ is a normalized factor and $\varrho_A={\rm Tr}_B(|\Phi\rangle_{AB}\langle\Phi|)$ denotes the reduced density operator of the subsystem $A$.

For a bipartite mixed state $\rho_{AB}$ on Hilbert space ${\cal H}_A\otimes {\cal H}_{B}$, the $S^{t}$-entropy entanglement is given via the convex-roof extension as
\begin{eqnarray}
E_t(\rho_{AB})=\inf_{\{p_i,|\Phi_i\rangle\}}\sum_ip_iE_t(|\Phi_i\rangle_{AB}),
\label{Eg2}
\end{eqnarray}
where $E_t(|\Phi_i\rangle_{AB})=S^{t}(\varrho^i_A)$  is the $S^{t}$-entropy of the subsystem $A$ for the pure state $\rho^i_{AB}=|\Phi_i\rangle_{AB}\langle\Phi_i|$, and the infimum is taken over all the possible pure-state decompositions of $\rho_{AB}=\sum_ip_i|\Phi_i\rangle\langle\Phi_i|$ with $p_i\geq0$, $\sum_ip_i=1$. The quantity $E_t(\rho)$ has defined an entanglement measure for any bipartite state.

\begin{theorem}
For an arbitrary finite-dimensional quantum state $\rho$ on Hilbert space ${\cal H}_A\otimes {\cal H}_{B}$, the quantity $E_t$ is a bipartite entanglement measure.

\end{theorem}

\emph{Proof}. For any density matrix $\rho$ on Hilbert space ${\cal H}_A\otimes {\cal H}_{B}$, it follows that $E_t(\rho)\geq 0$ from the non-negativity in Lemma 1 and Eq.(\ref{Eg2}), herein, the equality holds if and only if $\rho$ is separable. This has proved the condition (E1).

Note that the quantum $S^{t}$-entropy is invariant under local unitary transformations from Lemma 1, this follows the condition (E2).

The concavity of $S^{t}$-entropy in Lemma 1 assures the monotonicity under average local operations and classical communication (LOCC). Here, we firstly prove the monotonicity for pure states, i.e., when $\rho$ and $\sigma_j$ are all pure states. From the definition of the quantum $S^{t}$-entropy entanglement in Eq.(\ref{Eg1}), we have
\begin{eqnarray}
E_t(\rho)&=&S^{t}(\rho^A)
\nonumber
\\
&=&S^{t}(\sum_jp_j\sigma^A_j)
\nonumber
\\
&\geq& \sum_jp_jS^{t}(\sigma^A_j)
\label{LOCC1}
\\
&=&\sum_jp_jE_t(\sigma_j),
\label{LOCC2}
\end{eqnarray}
where the inequality (\ref{LOCC1}) is from the concavity of the quantum $S^{t}$-entropy in Lemma 1. The definition of the $S^{t}$-entropy entanglement in Eq.(\ref{Eg1}) implies Eq.(\ref{LOCC2}). According to Ref.\cite{Vidal2000}, the monotonicity for mixed states can be inherited from the monotonicity of pure states via the convex roof extension. This has proved the condition (E3).

Finally, the convexity (E4) is followed directly from the fact that all the entanglement measures constructed via the convex roof extension are convex \cite{3H2009}. $\Box$

\subsection{Analytic formula for two-qubit states}
\label{two-qubit}

In this subsection, we provide an analytic formula of  the $S^{t}$-entropy entanglement for two-qubit systems. For any bipartite pure state $|\Phi\rangle_{AB}$ on Hilbert space ${\cal H}_A\otimes {\cal H}_{B}$, its concurrence is defined  by \cite{Rungta2001}:
\begin{eqnarray}
C(|\Phi\rangle_{AB})=\sqrt{2(1-{\rm Tr}(\varrho^2_A))},
\label{eqC1}
\end{eqnarray}
where $\varrho_A={\rm Tr}_B(|\Phi\rangle_{AB}\langle\Phi|)$ is the reduced density operator of the subsystem $A$.

For a mixed state $\rho_{AB}$ on Hilbert space ${\cal H}_A\otimes {\cal H}_{B}$, the concurrence is given by the convex extension as
\begin{eqnarray}
C(\rho_{AB})=\inf_{\{p_i,|\Phi_i\rangle\}}
\sum_ip_iC(|\Phi_i\rangle_{AB}),
\label{eqC2}
\end{eqnarray}
where the infimum takes over all the possible pure-state decompositions of $\rho_{AB}$, i.e., $\rho_{AB}=\sum_ip_i|\Phi_i\rangle_{AB}\langle\Phi_i|$,  $\{p_i\}$ is a probability distribution with $p_i\geq 0$ and $\sum_ip_i=1$.

Interestingly, for a two-qubit mixed state $\rho$ on Hilbert space ${\cal H}_A\otimes {\cal H}_{B}$, its concurrence has the analytic formula as follows \cite{Coffman2000}:
\begin{eqnarray}
C(\rho)=\max\{0,\lambda_1-\lambda_2-\lambda_3-\lambda_4\},
\label{eqnC}
\end{eqnarray}
with $\lambda_i$'s being eigenvalues of the matrix $\sqrt{\rho(\sigma_Y\otimes\sigma_Y)\rho^* (\sigma_Y\otimes\sigma_Y)}$ in decreasing order, herein, $\rho^*$ is the complex conjugate of $\rho$, and $\sigma_Y$ denotes Pauli operator.

Now, consider any ${\cal C}^2\otimes {\cal C}^d$ pure state $|\phi\rangle_{AB}$ on Hilbert space $\mathcal{H}_A\otimes\mathcal{H}_B$ with Schmidt decomposition
\begin{eqnarray}
|\phi\rangle_{AB}=\sqrt{\lambda_0}|0\rangle|\phi_0\rangle+\sqrt{\lambda_1}|1\rangle|\phi_1\rangle,
\label{Schmidt}
\end{eqnarray}
 where the subsystem $A$ is qubit while the subsystem $B$ is on finite dimensional space, and $\{|\phi_i\rangle\}$ are orthogonal states on $\mathcal{H}_B$. From Eq.(\ref{Eg1}), we get the $S^{t}$-entropy entanglement as
\begin{eqnarray}
E_t(|\phi\rangle_{AB})=-\lambda_0\log_2\lambda_0-\lambda_1\log_2\lambda_1.
\label{hx1}
\end{eqnarray}
Besides, the concurrence of $|\Psi\rangle_{AB}$ is given by
\begin{eqnarray}
C(|\phi\rangle_{AB})=\sqrt{2(1-{\rm{Tr}}(\varrho^2_A))}=2\sqrt{\lambda_0\lambda_1}.
\label{C}
\end{eqnarray}
This can be proved that
\begin{eqnarray}
E_t(|\phi\rangle_{AB})=h(C(|\phi\rangle_{AB})),
\label{relation0}
\end{eqnarray}
where $h(x)$ is an analytic function defined as
\begin{eqnarray}
h(x)
&=&-\frac{1+\sqrt{1-x^2}}{2}\log_2\frac{1+\sqrt{1-x^2}}{2}
\nonumber\\
& &
-\frac{1-\sqrt{1-x^2}}{2}\log_2(\frac{1-\sqrt{1-x^2}}{2}).
\label{eqnhx}
\end{eqnarray}
Thus, we get a functional relation (\ref{relation0}) between the concurrence and $S^{t}$-entropy entanglement for any qubit-qudit pure state on Hilbert space $\mathcal{H}_A\otimes\mathcal{H}_B$.

In fact, for any two-qubit mixed state, the $S^{t}$-entropy entanglement has a similar relation with Eq.(\ref{relation0}). Here, we firstly prove the monotonicity and convexity of $h(x)$ in Eq.(\ref{eqnhx}).

\begin{proposition}
The function $h(x)$ defined in Eq.(\ref{eqnhx}) is monotonically increasing and convex.

\label{prop1}
\end{proposition}

\emph{Proof.} Since $h(x)$ is an analytic function on $x\in (0,1)$, its monotonicity and convexity can be followed from the nonnegativity of its first and second derivatives. In fact, the first derivative of $h(x)$ is given by
\begin{eqnarray}
\frac{d h(x)}{d x}=\frac{x}{2\sqrt{1-x^2}}\times\log_2\frac{1+\sqrt{1-x^2}}{1-\sqrt{1-x^2}}.
\label{dh1}
\end{eqnarray}
This is always nonnegative on $x\in[0,1]$. Moreover, we can get Eq. (\ref{dh1}) is positive for $x\in(0,1)$. This means that $h(x)$ is a strictly monotone increasing function on $x\in[0,1]$. While, from Eq.(\ref{dh1}) the second derivative of $h(x)$ is given by
\begin{eqnarray}
\frac{d^2h(x)}{dx^2}=\frac{-2\sqrt{1-x^2}+\ln(1+\sqrt{1-x^2})-\ln(1-\sqrt{1-x^2})}{2\ln2\times\sqrt{(1-x^2)^3}}.
\label{dh2}
\end{eqnarray}
which is positive for any $x\in(0,1)$.

For any given two-qubit mixed state $\rho$ there is an optimal pure-state decomposition $\rho_{AB}=\sum_ip_i|\Phi_i\rangle_{AB}\langle\Phi_i|$  such that its concurrence is equal to its average pure-state concurrences \cite{Wootters1998}, that is, we get
\begin{eqnarray}
C(\rho_{AB})=
\sum_ip_iC(|\Phi_i\rangle_{AB})
\label{eqnEC00}
\end{eqnarray}
This means that
\begin{eqnarray}
h(C(\rho_{AB}))
&=&h(\sum_ip_iC(|\Psi_{i}\rangle_{AB}))
\nonumber\\
&=&\sum_ip_ih(C(|\Psi_{i}\rangle_{AB}))
\label{eqnEC02}
\\
&=&\sum_ip_iE_t(|\Psi_{i}\rangle_{AB})
\label{eqnEC03}
\\
&\geq&E_t(\rho_{AB}),
\label{eqnEC2}
\end{eqnarray}
where the equality (\ref{eqnEC02}) is followed from Eq.(\ref{eqnEC00}), the equality (\ref{eqnEC03}) is obtained by using Eq.(\ref{relation0}), and the inequality (\ref{eqnEC2}) is due to the definition of $E_t(\rho_{AB})$ in Eq.(\ref{Eg2}).

Moreover, from the optimal decomposition of
$\rho_{AB}=\sum_jp_j|u_j\rangle\langle u_j|$ for the $S^{t}$-entropy entanglement we get that
\begin{eqnarray}
E_t(\rho_{AB})
\nonumber
&=&\sum_jp_jE_t(|u_j\rangle_{AB}))
\\
&=&\sum_jp_jh(C(|u_j\rangle_{AB}))
\label{eqnEC3}
\\
&\geq & h(\sum_jp_jC(|u_j\rangle_{AB}))
\label{eqnEC4}
\\
&\geq& h(C(\rho_{AB})),
\label{eqnEC5}
\end{eqnarray}
where the equality (\ref{eqnEC3}) is followed from the relation (\ref{relation0}), the  inequality (\ref{eqnEC4}) is derived from the convexity of $h(x)$ in Proposition 1, and the inequality (\ref{eqnEC5}) is from the monotonicity of $h(x)$ and the definition of $C(\rho_{AB})$ in Eq.(\ref{eqC2}).

Combing Eqs.(\ref{eqnEC2}) with (\ref{eqnEC5}), we have
\begin{eqnarray}
E_t(\rho_{AB})=h(C(\rho_{AB}))
\label{eqnEC6}
\end{eqnarray}
for any two-qubit mixed state $\rho_{AB}$. So, Eq.(\ref{eqnEC6}) provides an analytic formula of the $S^{t}$-entropy entanglement for any two-qubit state.

\section{ Monogamy inequality of the $S^{t}$-entropy entanglement in multi-dimensional systems}

\begin{figure}[htb]
\begin{center}
\resizebox{220pt}{80pt}{\includegraphics{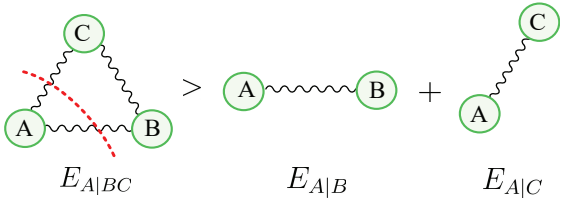}}
\end{center}
\caption{\small (Color online). A monogamy relation for tripartite systems $A$, $B$, and $C$ in terms of the entanglement measure $E$. Green circles represent the parties, wavy lines represent entangled sources, the red dotted curve represents the bipartition. }
\label{monogamy}
\end{figure}

The original monogamy of the entanglement is quantitatively shown in the inequality (\ref{eqn0}). For a tripartite system, its monogamy relation is shown in Fig.\ref{monogamy}. However, the inequality (\ref{eqn0}) may be invalid for general entanglement measures. Thus a natural question is to determine whether a given entanglement measure is monogamous or not.

\subsection{The $n$-qubit systems}

The $S^{t}$-entropy entanglement for qubit systems can be reduced to the entanglement of formation (EOF) corresponding to the case of $d=2$ in Eq.(\ref{hx1}). In fact, for a pure state $|\Phi\rangle_{AB}$ on Hilbert space ${\cal H}_A\otimes {\cal H}_{B}$, EOF is defined via the non Neumann entropy as \cite{Bennett19963824,Wootters1998}:
\begin{eqnarray}
E_f(|\Phi\rangle_{AB})=S(\varrho_A)=-{\rm{Tr}}(\varrho_A\log_2\varrho_A),
\label{eqn03}
\end{eqnarray}
where $\varrho_A={\rm Tr}_B(|\Phi\rangle_{AB}\langle\Phi|)$ denotes the reduced density operator of the subsystem $A$. EOF for a bipartite mixed state $\rho_{AB}$ on Hilbert space ${\cal H}_A\otimes {\cal H}_{B}$ is given by
\begin{eqnarray}
E_f(\rho_{AB})=\inf_{\{p_i,|\Phi_i\rangle\}}\sum_ip_iE(|\Phi_i\rangle)_{AB},
\end{eqnarray}
where the infimum takes over all the possible pure-state decompositions of $\rho_{AB}=\sum_ip_i|\Phi_i\rangle_{AB}\langle\Phi_i|$ and $\{p_i\}$ is a probability distribution. It has been proved that the squared EOF is monogamous for any $n$-qubit system $\rho_{A_1A_2\cdots A_n}$ on Hilbert space $\otimes_{i=1}^n\mathcal{H}_{A_i}$, that is,
\begin{eqnarray}
E^2_f(\rho_{A_1|A_2\cdots A_n})\geq \sum_{i=2}^nE^2_f(\rho_{A_1|A_i}).
\end{eqnarray}
where $E(\rho_{A_1|A_2\cdots A_n})$ characterizes the bipartite entanglement with respect to the bipartition $A_1$ and $A_2\cdots A_n$, and $E(\rho_{A_1|A_i})$ is the bipartite entanglement of the reduced density operator $\rho_{A_1|A_i}={\rm Tr}_{A_1\cdots A_{i-1}A_{i+1}\cdots A_{n}}(\rho_{A_1|A_2\cdots A_n})$ of joint subsystems $A_1$ and $A_i$ for $i=2, \cdots, n$.

From Eqs.(\ref{relation0}) and (\ref{eqnEC6}), there exists the same analytic  formula of
the $S^{t}$-entropy entanglement and EOF for any qubit system.  Combined with the monogamy properties of EOF \cite{Bai3,Bai2014,Zhu2014}, we can prove that the squared $S^{t}$-entropy entanglement is monogamous for any $n$-qubit system $\rho_{A_1A_2\cdots A_n}$ on Hilbert space $\otimes_{i=1}^n\mathcal{H}_{A_i}$, that is,
\begin{eqnarray}
E^2_t(\rho_{A_1|A_2\cdots A_n})\geq \sum_{i=2}^nE^2_t(\rho_{A_1|A_i}).
\end{eqnarray}
This means that both the EOF and $S^{t}$-entropy entanglement have the same monogamy features.

\subsection{High dimensional systems}

Our goal in this subsection is to extend the monogamy relationship for high dimensional systems. This will be explained by using several higher-dimensional states.

Consider a multipartite quantum system $|\Psi\rangle$ on Hilbert space $\otimes_{i=1}^n\mathcal{H}_{A_i}$. Define the ``residual entanglement'' of the $S^{t}$-entropy entanglement as
\begin{eqnarray}
\tau_\gamma^{E_t}(|\Psi\rangle_{A_1A_2\cdots A_n})=E_t^\gamma(|\Psi\rangle_{A_1|A_2\cdots A_n})-\sum^n_{i=2}E_t^\gamma(\rho_{A_1|A_i}),
\end{eqnarray}
where $E_t(|\Psi\rangle_{A_1|A_2\cdots A_n})$ denotes the bipartite entanglement with respect to the bipartition $A_1$ and $A_2\cdots A_n$, and $E_t(\rho_{A_1|A_i})$ is the bipartite entanglement of the reduced density operator $\rho_{A_1|A_i}$ of joint subsystems $A_1$ and $A_i$ for $i=2, \cdots, n$, and $\gamma$ is a parameter with $\gamma\in (0,\infty)$. We show that the $S^{t}$-entropy entanglement has different monogamy relationship from EOF.

\begin{proposition}
\label{SgEOF}
The monogamy of $S^{t}$-entropy entanglement and EOF are inequivalent for high dimensional entangled systems.

\end{proposition}

Proposition \ref{SgEOF} is proved by using the following examples.

\emph{Example 3.} Consider a $4 \otimes 2 \otimes 2$ system \cite{Luo2022} in the pure state $|\Psi\rangle_{ABC}$ given by
\begin{eqnarray}
|\Psi\rangle_{ABC}=\frac{1}{\sqrt{2}}(\alpha|000\rangle+\beta|110\rangle+\alpha|201\rangle+\beta|311\rangle), \label{chain001}
\end{eqnarray}
where $\alpha$ and $\beta$ satisfies that $\alpha^2+\beta^2=1$. The reduced density operator of the subsystem $A$ is given by
\begin{eqnarray}
\varrho_A&=&\frac{\alpha^2}{2}|00\rangle\langle00|+\frac{\beta^2}{2}|01\rangle\langle01|.
\nonumber
\\
&&+\frac{\alpha^2}{2}|10\rangle\langle10|+\frac{\beta^2}{2}|11\rangle\langle11|.
\end{eqnarray}

From Eq.(\ref{Eg1}) we get
\begin{eqnarray}
E_t(|\Psi\rangle_{A|BC})=\frac{a+b+4}{8-3\log_23},
\label{3tripartite}
\end{eqnarray}
where $a=-\alpha^2\log_2\alpha^2-(2-\alpha^2)\log_2(2-\alpha^2)$ and $b=-\beta^2\log_2\beta^2-(2-\beta^2)\log_2(2-\beta^2)$. The reduced state for the joint system $A$ and $B$ is given by
\begin{eqnarray}
\rho_{AB}=\frac{1}{2}(|\psi_1\rangle\langle \psi_1|+|\psi_2\rangle\langle \psi_2|),
\end{eqnarray}
where $|\phi_i\rangle$ are defined by $|\psi_1\rangle=\alpha|00\rangle+\beta|11\rangle)$ and $|\psi_2\rangle=\alpha|20\rangle+\beta|31\rangle$. For any pure state decomposition of $\rho_{AB}$, the pure state component has the form
\begin{eqnarray}
|\mu_i\rangle=a_i|\psi_1\rangle+e^{-i\delta}\sqrt{1-a^2_i}|\psi_2\rangle,
\end{eqnarray}
In this case, the reduced density operator is given by $\varrho^i_{B}=\alpha^2|0\rangle\langle0|+\beta^2|1\rangle\langle1|$. Therefore, according to the definition of the $S^{t}$-entropy entanglement in Eq.(\ref{Eg2}), we have
\begin{eqnarray}
E_t(\rho_{AB})=\frac{-2\alpha^2\log_2\alpha^2-2\beta^2\log_2\beta^2}{8-3\log_23}.
\end{eqnarray}

\begin{figure}[htb]
\begin{center}
\resizebox{240pt}{120pt}{\includegraphics{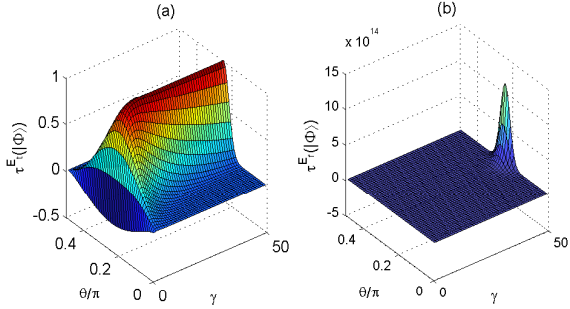}}
\end{center}
\caption{\small (Color online). The indicator $\tau$ for a $4 \otimes2 \otimes 2$ dimensional entanglement in Example 3. (a) The $S^{t}$-entropy entanglement. (b) The EOF entanglement. Here, $\alpha=\cos\theta$ and $\beta=\sin\theta$. $\tau^{E_t}$ is defined in Eq.(\ref{Et}) while $\tau^{E_f}$ is defined in Eq.(\ref{Ef}).}
\label{EgEf}
\end{figure}

In a similar manner, for the reduced quantum state $\rho_{AC}$, the reduced density operator is given by $\varrho^i_{C}=\frac{1}{2}|0\rangle\langle0|+\frac{1}{2}|1\rangle\langle1|$. This implies that
\begin{eqnarray}
E_t(\rho_{AC})=\frac{2}{8-3\log_23}.
\end{eqnarray}
So, the ``residual tangle'' of the $S^{t}$-entropy entanglement is given by
\begin{eqnarray}
 \tau^{E_t}(|\Psi\rangle_{A|BC})=E_t(|\Psi\rangle_{A|BC})-E_t(\rho_{AB})
 -E_t(\rho_{AC})\not=0
 \label{Et}
\end{eqnarray}
for $\gamma=1$. Instead, the ``residual tangle'' of the EOF entanglement is given by \cite{Bai2014}:
\begin{eqnarray}
&&E_f(|\Psi\rangle_{A|BC})=-\alpha^2\log_2\alpha^2-\beta^2\log_2\beta^2+1
\\
&&E_f(\rho_{AB})=-\alpha^2\log_2\alpha^2-\beta^2\log_2\beta^2
\\
&&E_f(\rho_{AC})=1
\end{eqnarray}
which follows
\begin{eqnarray}
\tau_\gamma^{E_f}(|\Psi\rangle_{A|BC})=E_f^\gamma(|\Psi\rangle_{A|BC})-E_f^\gamma(\rho_{AB})
-E_f^\gamma(\rho_{AC})=0
\label{Ef}
\end{eqnarray}
for $\gamma=1$. For general $\gamma\not=1$, both residual tangles are shown Fig.\ref{EgEf}. We obtain the residual tangle of $\tau^{E_t}$ is negative while $\tau^{E_f}$ is zero.

\emph{Example 4}. Consider a $6\otimes3\otimes3$ dimensional system in the pure state $|\Phi\rangle_{ABC}$ given by
\begin{eqnarray}
|\Phi\rangle_{ABC}&=&\frac{1}{2\sqrt{3}}(|012\rangle+|021\rangle+|120\rangle+102\rangle
+|210\rangle
\nonumber\\
&&
 +|201\rangle)+\frac{1}{\sqrt{6}}(|300\rangle+|411\rangle+|522\rangle).
\label{high4}
\end{eqnarray}
The reduced density operator of the subsystem $A$ is given by $\varrho_A=\frac{1}{6}\mathbbm{1}$ with the identity matrix $\mathbbm{1}$. According to Eq.(\ref{Eg1}) we have
\begin{eqnarray}
E_t(|\Phi\rangle_{A|BC})=1.
\label{Example400}
\end{eqnarray}

Note that the reduced density operator $\rho_{AB}$ of $|\Psi\rangle_{ABC}$ in Eq.(\ref{high4}) has the following spectral decomposition
\begin{eqnarray}
\rho_{AB}&=&\frac{1}{3}(|x_1\rangle\langle x_1|+|x_2\rangle\langle x_2|+|x_3\rangle\langle x_3|),
\label{highAB}
\end{eqnarray}
where $|x_i\rangle$ are eigenstates given by
\begin{eqnarray}
&&|x_1\rangle_{AB}=\frac{1}{2}|21\rangle+\frac{1}{2}|12\rangle+\frac{1}{\sqrt{2}}|30\rangle), \nonumber\\
&&|x_2\rangle_{AB}=\frac{1}{2}|02\rangle+\frac{1}{2}|20\rangle+\frac{1}{\sqrt{2}}|41\rangle),
\nonumber\\
&&|x_3\rangle_{AB}=\frac{1}{2}|10\rangle+\frac{1}{2}|01\rangle+\frac{1}{\sqrt{2}}|52\rangle).
\label{}
\end{eqnarray}
 According to the Hughston-Jozsa-Wootters (HJW) Theorem \cite{Hughston1993}, any pure state ensemble of $\rho_{AB}$ can be expressed as a superposition of $|x_i\rangle$ with $i=1,2, 3$, namely, for arbitrary pure state $|\varphi_i\rangle_{AB}=\sum_ic_i|x_i\rangle$ with $\sum_ic^2_i=1$, its reduced density operator $\varrho_A={\rm Tr}_{B}(|\varphi_i\rangle_{AB}\langle\varphi_i|)$ has the same spectrum $\{\frac{1}{2},\frac{1}{4},\frac{1}{4}\}$. So, we have
\begin{eqnarray}
E_t(|\varphi_i\rangle_{AB})=\frac{5-\frac{3}{2}\log_23}{3\log_23-2}=0.9520.
\label{phi}
\end{eqnarray}
Moreover, the state $\rho_{AB}$ in Eq. (\ref{highAB}) can be decomposed into
\begin{eqnarray}
\rho_{AB}=\sum_ip_i|\varphi_i\rangle_{AB}\langle\varphi_i|.
\end{eqnarray}
Thus, we have
\begin{eqnarray}
E_t(\rho_{AB})=\sum_ip_iE_t(|\varphi_i\rangle_{AB})= 0.9520
\label{Example40}
\end{eqnarray}
from Eq.(\ref{phi}).

Similarly, we show that the EOF of $\rho_{AC}$ is given by
\begin{eqnarray}
E_t(\rho_{AC})= 0.9520.
\label{Example41}
\end{eqnarray}
From Eqs.(\ref{Example400}), (\ref{Example40}), and (\ref{Example41}), we have
\begin{eqnarray}
E_t^{\alpha}(|\Phi\rangle_{A|BC})<E_t^{\alpha}(\rho_{AB})+E^{\alpha}_t(\rho_{AC}).
\label{Example42}
\end{eqnarray}
for any $\alpha\leq 14$, or
\begin{eqnarray}
E_t^{\alpha}(|\Phi\rangle_{A|BC})>E_t^{\alpha}(\rho_{AB})+E_t^{\alpha}(\rho_{AC}).
\label{Example42}
\end{eqnarray}
for any $\alpha\geq 15$, which is going beyond the monogamy inequality (\ref{eqn0}) with the squared EOF \cite{Oliveira2014,Bai3,Bai2014},  the squared R\'{e}nyi-$\alpha$ entropy \cite{R2015}, the squared Tsallis-$q$ entropy \cite{Luo2016}, and the squared Unified-$(r, s)$ entropy \cite{Khan2019}.

For the EOF, we have
\begin{eqnarray}
&& E_f(|\Phi\rangle_{A|BC})=\log_26,
\nonumber\\
&&E_f(\rho_{AB})=E_f(\rho_{AC})=\frac{3}{2}.
\label{Example43}
\end{eqnarray}
From Eq.(\ref{Example43}) it is clearly that
\begin{eqnarray}
E^2_f(|\Phi\rangle_{A|BC})>E^2_f(\rho_{AB})+E^2_f(\rho_{AC}).
\label{Example44}
\end{eqnarray}
Consequently, Eqs.(\ref{Example42}) and (\ref{Example44}) has shown the inequivalence of the $S^{t}$-entropy entanglement and EOF in Proposition \ref{SgEOF}.

In a similar manner, using the total entropy of Tsallis entropy and its complementary dual, we can define a class of one-parameter entanglement measure, $\mathcal{T}^{t}_q$-entropy entanglement, as stated in  Appendix B, it also includes the monogamy properties of $\mathcal{T}^{t}_q$-entropy entanglement for qubit systems or higher-dimensional systems.

\section{Entanglement polygon inequalities for quantum networks}

\begin{figure}
\begin{center}
\resizebox{130pt}{120pt}{\includegraphics{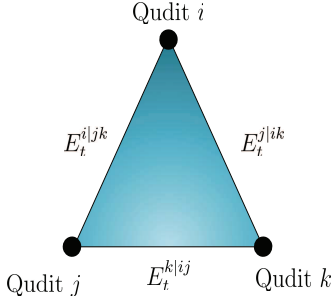}}
\end{center}
\caption{\small (Color online) Schematic polygon inequality (\ref{eqn00}) of tripartite entangled systems. The length of each side denotes correspondingly to the value of the marginal entanglement.}
\label{triangle}
\end{figure}

The monogamy inequality (\ref{eqn0}) provides a lower bound for the corresponding ``one-to-group'' entanglement, while the polygon inequality provides its upper bound. The polygon inequalities hold for arbitrary qubit pure states with respect to generic entanglement measures \cite{Qian2018}. A natural problem is whether the polygon inequalities can be generalized for higher-dimensional systems. A simple geometric interpretation for the inequality (\ref{eqn00}) is $E_{t}^{j|ik}(|\psi\rangle), E_{t}^{k|ij}(|\psi\rangle)$ and $E_{t}^{i|jk}(|\psi\rangle)$ consist of a triangle, that is, the one-to-group marginal entanglement $E_{t}$ can denote the lengths of the three edges of a triangle, as illustrated in Fig.\ref{triangle}.

\begin{figure}[ht]
\begin{center}
\resizebox{220pt}{110pt}{\includegraphics{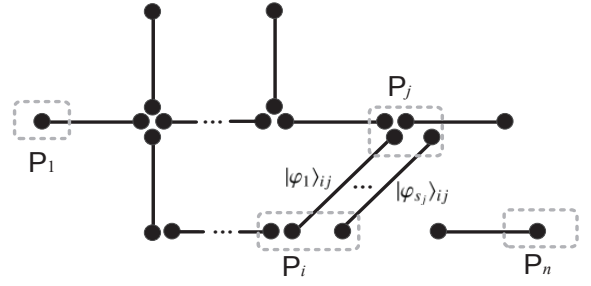}}
\end{center}
\caption{\small (Color online). An $n$-partite quantum network consisting of $n$ parties $P_1, \cdots, P_n$. Here, each pair of two parties $P_i$ and $P_j$ share $s_j$ pairs of bipartite entangled pure states $|\varphi_{1}\rangle_{ij}, |\varphi_{2}\rangle_{ij},\cdots, |\varphi_{s_j}\rangle_{ij}$.}
\label{Fig5a}
\end{figure}

One special scenario to characterize general high-dimensional entangled states is from distributive constructions, where the local tensor structures further can be regarded as quantum networks. In fact, consider an $n$-partite entangled quantum network $\mathcal{N}_q(\cal{P},{\cal E})$ consisting of bipartite entangled pure states a shown in Fig.\ref{Fig5a}, where $\cal{P}$ represents parties $P_1, \cdots, P_n$, and ${\cal E}$ denotes entangled states. Suppose that there are $s_j$ pairs of arbitrary bipartite entangled pure states $|\varphi_{1}\rangle_{ij}, |\varphi_{2}\rangle_{ij},\cdots, |\varphi_{s_j}\rangle_{ij}$ shared by any two parties $P_i$ and $P_j$. Then, the joint state shared by two parties $P_i$ and $P_j$ is given by $\mathop{\otimes}^{s_j}_{s=1}|\varphi_{s}\rangle_{ij}$, the total state of $\mathcal{N}_{q}(\cal{P},{\cal E})$ is denoted as
\begin{eqnarray}
 |\Psi\rangle_{P_1\cdots P_n}
 =\mathop{\otimes}^{n}_{1\leq i<j\leq n}\mathop{\otimes}^{s_j}_{s=1}|\varphi^{s}\rangle_{ij}.
\end{eqnarray}
Next, we prove the polygon inequality for high-dimensional systems derived from quantum network $\mathcal{N}_{q}$.

\begin{theorem}
\label{marginal01}
The polygon inequality of the quantum network $\mathcal{N}_q(\cal{P},{\cal E})$ is given by
\begin{eqnarray}
E^{i|\overline{i}}_{t}(|\Psi\rangle)\leq \sum_{j\neq i}E^{j|\overline{j}}_{t}(|\Psi\rangle),
\end{eqnarray}
where $E^{i|\overline{i}}_t$ denotes the one-to-group entanglement of the $n$-qudit quantum network state $|\Psi\rangle$, the subscript $i$ refers to all the systems owned by the party $P_i$, and $\overline{i}$ denotes all the systems except for its owned by $P_i$, and $E_t$ is the $S^{t}$-entropy entanglement.

\end{theorem}

\emph{Proof.} For a given party $P_i$ in the quantum network $\mathcal{N}_{q}$, assume that it has entangled with $N$ parties $P_j$'s. Instead, suppose there are $M$ parties $P_k$'s who are entangled with the party $P_j$.  It is sufficient to prove that
\begin{eqnarray}
E^{i|\overline{i}}_{t}(|\Psi\rangle)
\nonumber
&=&
E_t(\mathop{\otimes}^{N}_{j=1}\mathop{\otimes}^{s_j}_{s=1}|\varphi^{s}\rangle_{ij})
\nonumber \\
&=&S(\otimes_j\rho^j_i)
\label{network0}\\
&=&S(\otimes_i\rho^i_j)
\label{network1}
\\
&\leq&\sum_{j\neq i}S(\otimes_k\rho^k_j)
\label{network2}
\\
&=&\sum_{j\neq i}E^{j|\overline{j}}_{t}(|\Psi\rangle),
\label{network3}
\end{eqnarray}
where the state $\rho^j_i={\rm Tr}_i(|\psi\rangle_{ij}\langle\psi\rangle_{ij}|)$ in Eq.(\ref{network0}) is the reduced density operator for the subsystem of $P_i$ by tracing out the subsystems of $P_j$ entangled with  $P_i$ and $|\psi\rangle_{ij}=\mathop{\otimes}^{s_j}_{s=1}|\varphi_{s}\rangle_{ij}$ is the total state  entangled between $P_i$  and $P_j$. The equality (\ref{network1}) is due to the symmetry in Lemma 1, where the state $\rho^i_j={\rm Tr}_j(|\psi\rangle_{ij}\langle\psi\rangle_{ij}|)$ is the reduced density operator of the subsystems of $P_j$. The inequality (\ref{network2}) follows from the additivity of separable state $\rho^k_j$ in Eq.(\ref{subadditivity}) of Lemma 1, herein, $\rho^k_j$ is the reduced density operator for the subsystem of $P_j$ by tracing out the subsystems of all the parties $P_k$. Here, $\{P_k\}$ denotes all the parties entangled with the given party $P_j$, i.e., the number of $\{P_i\}$ is no more than the numbers of $\{P_k\}$. $\Box$

We define the indicator $\tau$ of the $S^{t}$-entropy marginal entanglement for multipartite quantum system $|\Psi\rangle_{P_1\cdots P_n}$ as
\begin{eqnarray}
\tau^{E_t}(|\Psi\rangle_{P_1\cdots P_n})=E^{i|\overline{i}}_{t}(|\Psi\rangle)-\sum_{j\neq i,\forall j}E^{j|\overline{j}}_{t}(|\Psi\rangle).
\label{indicator1}
\end{eqnarray}

\emph{Example 5}. Consider a $3$-partite chain quantum network state $|\Psi\rangle_{ABC}$ in Example 3.

\begin{figure}[htb]
\begin{center}
\resizebox{220pt}{180pt}{\includegraphics{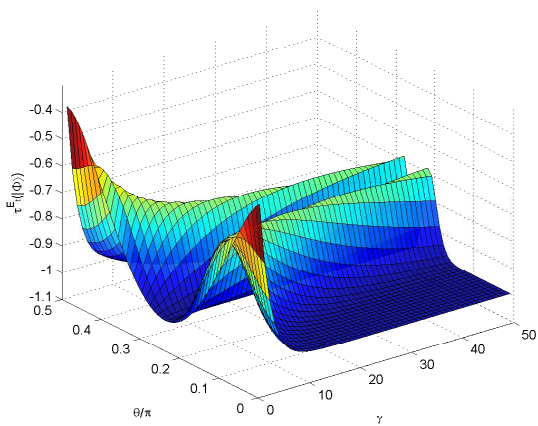}}
\end{center}
\caption{\small (Color online). The indicator $\tau_\gamma^{E_t}$ of $4 \otimes2 \otimes 2$ dimensional system in Example 6. Here, $a=\cos\theta$ and $b=\sin\theta$.}
\label{EPI(networkstate)1}
\end{figure}

The reduced density operator of the subsystem $A$, $B$, and $C$ is respectively given by
\begin{eqnarray}
&&
\varrho_A=\frac{\alpha^2}{2}|0\rangle\langle 0|+\frac{\beta^2}{2}|1\rangle\langle 1|
+\frac{\alpha^2}{2}|2\rangle\langle 2|+\frac{\beta^2}{2}|3\rangle\langle 3|,
\nonumber\\
&&\varrho_B=\alpha^2|0\rangle\langle 0|+\beta^2|1\rangle\langle 1|,
\nonumber\\
&&\varrho_C=\frac{1}{2}|0\rangle\langle 0|+\frac{1}{2}|1\rangle\langle 1|.
\end{eqnarray}
From Eq.(\ref{Eg1}), we get
\begin{eqnarray}
&&E^{A|BC}_t(|\Psi\rangle)=\frac{a+b+4}{8-3\log_23},
\nonumber \\
&&E^{B|AC}_t(|\Psi\rangle)=-\alpha^2\log_2\alpha^2-\beta^2\log_2\beta^2,
\nonumber \\
&& E^{C|AB}_t(|\Psi\rangle)=1,
\label{marginal0}
\end{eqnarray}
where $a$, $b$ are given in Eq.(\ref{3tripartite}). According to Eq.(\ref{indicator1}) we get
\begin{eqnarray}
\tau^{E_t}(|\Psi\rangle)
&=&E^{A|BC}_t(|\Psi\rangle)-E^{B|AC}_t(|\Psi\rangle)-E^{C|AB}_t(|\Psi\rangle).
\label{marginal20}
\end{eqnarray}
Combining with Eqs.(\ref{marginal0}) and (\ref{marginal20}), the indicator $\tau$ of the $S^{t}$-entropy marginal entanglement is shown in Fig.\ref{EPI(networkstate)1}. It states that $\tau^{E_t}(|\Psi\rangle)$ is negative. This shows Theorem \ref{marginal01} holds for quantum network state defined in Eq.(\ref{chain001}).

\section{Conclusion}

Entanglement entropy exhibits qualitatively different behaviours from that of classical entropy, and provides a valuable tool for describing the features of many-body states. More importantly, the entanglement entropy measures how a subsystem is entangled with other systems in a given system. In an operational manner, the degree of entanglement in a pure many-body state is quantified by the entanglement entropy of its subsystems.  However, until now, most of the entanglement entropies ignore the information of its complement of subsystems. Our consideration in this paper is to propose a new measure from non Neumann entropy and its complement entropy to characterize entanglement. This kind of entanglement shows new features beyond previous entanglement measures. So far, we have considered some special entanglement. A natural problem is to investigate the monogamy relationship for general entanglements, specially for high-dimensional systems.

In conclusion, we proposed the total entropy of the von Neumann entropy and its  complementary dual which not only reveals the information of one subsystem but also its complementary statistics. We defined the bipartite $S^{t}$-entropy entanglement based on the total entropy and provided the analytic formula in two-qubit systems. We also considered the monogamy property of qubit systems and higher-dimensional systems in terms of the $S^{t}$-entropy entanglement. We obtained a polygon monogamy inequality for arbitrary quantum networks consisting of bipartite entangled pure states. This reveals an intriguing feature of high-dimensional entanglement going beyond any qubit entanglement. These results are interesting in quantum entanglement theory.

\section*{Acknowledgments}

This work was supported by the National Natural Science Foundation of China (Nos.62172341,61772437), Fundamental Research Funds for the Central Universities (No.2018GF07), and Shenzhen Institute for Quantum Science and Engineering.



\appendix

\section{Proof of Lemma 1}

The proof is completed by one-by-one.

(i) The quantum total entropy $S^{t}$ for any density matrix $\rho$ on $d$-dimensional Hilbert space $\mathcal{H}$ satisfies the following inequalities as
\begin{eqnarray}
0\leq S^{t}(\rho)\leq d\log_2d-(d-1)\log_2(d-1).
\label{}
\end{eqnarray}
Here, the lower bound is achieved only for pure states while the upper bound is reached for the maximally mixed state $\rho=\frac{\mathbb{I}}{d}$.

Define a function $g: [0,1]\mapsto \mathbb{R}$ as
\begin{eqnarray}
g(x)=-x\log_2x-(1-x)\log_2(1-x).
\label{gx}
\end{eqnarray}
It is easy to calculate the second derivative of $g(x)$ as
\begin{eqnarray}
\frac{d^2g(x)}{dx^2}=-\frac{1}{x(1-x)\ln2}<0.
\label{gx0}
\end{eqnarray}
This implies that $g(x)$ is a concave function.

From Eq.(\ref{entropy1}) it follows that
\begin{eqnarray}
S^{t}(\rho)&=&{\rm Tr}[g(\rho)]
\nonumber
\\
&=&\sum_{i}g(\lambda_i)
\nonumber
\\
&=&d\sum_{i}\frac{1}{d}g(\lambda_i)
\nonumber
\\
&\leq& d g(\sum_{i}\frac{1}{d}\lambda_i)
\label{concavity0}
\\
&=&dg(\frac{1}{d})
\label{concavity1}
\\
&=&d\log_2d-(d-1)\log_2(d-1).
\label{concavity2}
\end{eqnarray}
Here, the inequality (\ref{concavity0}) is obtained from the concavity of the function $g(x)$ defined in Eq.(\ref{gx}). The equality (\ref{concavity1}) is followed from the equality of $\sum_{i}\lambda_i=1$. The equality (\ref{concavity2}) is derived from the definition of $g(x)$ in Eq.(\ref{gx}). This has proved the non-negativity of the total quantum entropy.

(ii)  For any convex combination $\rho=\sum^n_ip_i\rho_i$ on finite dimensional Hilbert space $\mathcal{H}_A\otimes \mathcal{H}_B$, we have
\begin{eqnarray}
\sum^n_ip_iS^{t}(\rho_i)\leq S^{t}(\rho),
\label{}
\end{eqnarray}
where $\{p_i\}$ is a probability distribution, and $\rho_i$'s are density matrices on finite dimensional Hilbert space $\mathcal{H}_A\otimes \mathcal{H}_B$. The equality holds if and only if all $\rho_i$'s are equal to each other. In fact, for any concave function $g(x)$, the functional $p=\rm{Tr}(g(\rho))$ is concave for any arbitrary Hermitian operator $\rho$ (see Sec.III in Ref. \cite{Rastegin2012} for details). According to Eq.(\ref{gx}), $g(x)$ is concave, that is, the concavity of $g$ ensures the concavity of $S^{t}(\rho)$. This has proved the concavity.

(iii) Suppose a pure state $|\psi\rangle_{AB}$ defined on Hilbert space ${\cal H}_A\otimes {\cal H}_{B}$ has the Schmidt decomposition
\begin{eqnarray}
|\psi\rangle=\sum^m_{i=1}\sqrt{\lambda_i}|a_i\rangle_A|b_i\rangle_B.
\label{eqnSchmidt1}
\end{eqnarray}
We get that the reduced density matrices $\varrho_A$ and $\varrho_B$ have the same spectra as $\{\lambda_i\}$. From Eq.(\ref{entropy1}) it follows that $S^{t}(\varrho_A)=S^{t}(\varrho_B)$ which has proved the symmetry.

(iv) Suppose a mixed state $\rho$ defined on Hilbert space ${\cal H}_{A}$ has a diagonal form:
\begin{eqnarray}
\rho=\sum^m_{i=1}\sqrt{\lambda_i}|e_i\rangle\langle e_i|,
\label{}
\end{eqnarray}
where $|e_i\rangle$ are orthogonal states. For a given unitary transformation $U\in \mathbb{SU}({\cal H})$, we have $U\rho U^\dag=\sum^m_{i=1}\sqrt{\lambda_i}|f_i\rangle\langle f_i|$, where $|f_i\rangle:=U|e_i\rangle$ are orthogonal states. Note that both $\rho$ and $U\rho U^\dag$ have the same eigenvalues. Thus from the definition of $S^{t}$-entropy in Eq.(\ref{entropy1}) we have concluded the invariance under the unitary operation.

(v) From Ref.\cite{Canosa2002} any general entropy function (\ref{general0}) is subadditive if $x\frac{d^2g(x)}{dx^2}$ is a strictly decreasing function of $x\in (0,1)$. From this condition, we can prove the subadditivity of the total quantum entropy $S^{t}$. In fact, it is easy to check that
\begin{eqnarray}
\frac{d xg''(x)}{dx}=-\frac{1}{(1-x)^2\ln2}<0,
\label{}
\end{eqnarray}
where $g''(x)=\frac{d^2g(x)}{dx^2}$ with $g(x)$ defined in Eq.(\ref{gx}). This implies that the quantum entropy $S^{t}$ is subadditive satisfying Eq.(\ref{subadditivity}). Moreover, we can obtain Eq.(\ref{trace0}) from Proposition 14 in Ref. \cite{Bosyk2016}.

(vi) Consider a given mixed state $\rho=\sum_ip_i|\psi_i\rangle\langle\psi_i|$, where $\{p_i\}$ is a probability distribution with $\sum_ip_i=1$ and $\{|\psi_i\rangle\}$ are pure states (orthogonal or nonorthogonal). Then our goal here is to prove
\begin{eqnarray}
S^{t}(\rho)\leq H^{t}(X),
\label{bound0}
\end{eqnarray}
where $H^{t}(X)$ is the classical total entropy defined as
\begin{eqnarray}
H^{t}(X)=-\sum^n_{i=1}(p_i\log_2p_i+(1-p_i)\log_2(1-p_i)),
\label{bound0}
\end{eqnarray}
and $X$ is the random variable associated with distribution $\{p_1, p_2, \cdots, p_n\}$. In fact, denote $\lambda_j$ and $|\varphi_j\rangle$ as the eigenvalues and the corresponding eigenstate of $\rho=\sum_j\lambda_j|\varphi_j\rangle\langle\varphi_j|$. The ensemble classification theorem \cite{Hughston1993} states that
\begin{eqnarray}
|\psi_i\rangle=\sum_ju_{ij}\sqrt{\lambda_j}|\varphi_j\rangle
\label{u0}
\end{eqnarray}
up to a normalization constant, where $(u_{ij})$ is a proper unitary matrix. It follows from Eq.(\ref{u0}) and $\langle\varphi_j|\varphi_k\rangle=\delta_{jk}$, $p_i=\sum_jw_{ij}\lambda_j$, where $w_{ij}=u^{\ast}_{ij}u_{ij}$ are elements of a matrix, i.e., $\sum_jw_{ij}=\sum_iw_{ij}=1$ for any $i,j$. Define the function $f(x)=-g(x)$, that is, $f: x\mapsto -g(x)\in \mathbb{R} $. $f(x)$ is convex function from the concavity of $g(x)$. Hence, we obtain
\begin{eqnarray}
\sum_if(p_i)\nonumber&=&\sum_if(\sum_jw_{ij}\lambda_j)
\\ &\leq& \sum_i\sum_jw_{ij}f(\lambda_j)
\label{Jensen00}
\\&=&\sum_jf(\lambda_j),
\label{Jensen0}
\end{eqnarray}
where the inequality (\ref{Jensen00}) is followed from the convexity of $f(x)$, the equality is due to $\sum_iw_{ij}=1$ for all $j$. Combined Eqs.(\ref{entropyc1}) with (\ref{entropy1}), the inequality (\ref{Jensen0}) implies the inequality (\ref{bound0}).

\section{The measure based the total entropy of Tsallis entropy and its complementary dual}

For the probability distribution $\textbf{p}=\{p_1,p_2,\cdots,p_n\}$ of random variable $X$, the Tsallis entropy \cite{Tsallis1988} is defined as
\begin{eqnarray}
T_q({\textbf{p}})=-\sum^n_{i=1}p_i\ln_qp_i
\end{eqnarray}
with one parameter $q$ as an extension of shannon entropy. This relation gives the arithmetic mean of the information $-\ln_qp_i$, with its corresponding probability $p_i$, where $q$ logarithm function is defined by
\begin{eqnarray}
\ln_q(x)=\frac{1-x^{1-q}}{q-1}
\end{eqnarray}
for any negative real numbers $x$ and $q$. The complementary dual of the Tsallis entropy is defined as
\begin{eqnarray}
\overline{T}_q({\textbf{p}})=-\sum^n_{i=1}(1-p_i)\ln_q(1-p_i).
\end{eqnarray}
Similar with Eq.(\ref{entropyc1}), we define the total entropy of Tsallis entropy and its complementary dual as
\begin{eqnarray}
T^{t}_q({\textbf{p}})=\sum^n_{i=1}T_q(p_i,1-p_i),
\end{eqnarray}
where $T_q(p_i,1-p_i)$ are the sum of the Tsallis entropy and its complementary dual of a discrete random variable $X$ with a probability distribution $\{p_1, p_2,\cdots,p_n\}$. $T^{t}_q$ is also viewed as entropy of  the binomial random variable $X_i\sim\{p_i,1-p_i\}$ with a probability distribution $\{p_i,1-p_i\}$.

The quantum entropy of a density operator equals the classical entropy of
the probability vector formed by its eigenvalues, thus we give the following definition.

\begin{definition}
The total entropy of the Tsallis entropy and its complementary dual of a quantum state $\rho$ on $d$-dimensional Hilbert space ${\cal H}$ is defined by
\begin{eqnarray}
T^{t}_q(\rho)&=&\frac{1-{\rm Tr}\rho^q-{\rm Tr}({\mathbbm{1}-\rho})^q+{\rm Tr}({\mathbbm{1}-\rho)}}{q-1},
\label{}
\end{eqnarray}
which corresponds to with the form (\ref{general0}) with
\begin{eqnarray}
f(p)=\frac{(p-p^q)+1-p-(1-p)^q}{(q-1)}.
\label{tp}
\end{eqnarray}
\end{definition}
Here, the function $f(p)$ is a smooth and strictly concave real function on $p\in [0,1]$ satisfying $f(0)=f(1)=0$. For the second derivative of $f(p)$, we have
\begin{eqnarray}
f''(p)=-q[(1-x)^{q-2}+p^{q-2}]<0
\end{eqnarray}
for any $q>0$. From Ref. \cite{Canosa2002}, this entropy satisfies most basic properties of the conventional entropy except for the additivity in Lemma 1.

For a pure state $|\Phi\rangle_{AB}$ on Hilbert space ${\cal H}_A\otimes {\cal H}_{B}$, we define its $T^{t}_q$-entropy entanglement as
\begin{eqnarray}
\mathcal{T}^{t}_q(|\Phi\rangle_{AB})=T^t_q(\varrho_A),
\label{Tq1}
\end{eqnarray}
where $\varrho_A={\rm Tr}_B(|\Phi\rangle_{AB}\langle\Phi|)$ is given in Eq.().

For a bipartite mixed state $\rho_{AB}$ on Hilbert space ${\cal H}_A\otimes {\cal H}_{B}$, the $T^{t}_q$-entropy entanglement is defined via convex-roof extension
\begin{eqnarray}
\mathcal{T}^{t}_q(\rho_{AB})=\inf_{\{p_i,|\Phi_i\rangle\}}\sum_ip_i\mathcal{T}^{t}_q(|\Phi_i\rangle_{AB}),
\label{Tq2}
\end{eqnarray}
where the infimum is taken over all the possible pure-state decompositions of $\rho_{AB}=\sum_ip_i|\Phi_i\rangle_{AB}\langle\Phi_i|$.

In fact, Since $T^{t}_q$-entropy complies the general entropic forms,  it follows from Proposition 2 in Ref.\cite{Canosa2002} that $\mathcal{T}^{t}_q(\rho)\geq 0$ for any density matrix $\rho$, where the equality holds if and only if $\rho$ is separable.
On the other hand, the $T^{t}_q$-entropy is invariant under local unitary transformation from Proposition 6 in Ref. \cite{Canosa2002}. This implies the condition (E2). The  concavity of the entropy $T^{t}_q(\rho)$ can be followed from the concavity of the function $f(p)$ in Eq.(\ref{tp}). As a result, the concavity of the $T^{t}_q$-entropy  ensures the monotonicity of $T^{t}_q$ under average LOCC, that is, the condition (E3) holds. The convexity (E4) is followed directly from the fact that all the measures constructed via the convex roof extension are convex \cite{3H2009}.

Consider an arbitrary pure state $\phi_{AB}$ given in Eq.(\ref{Schmidt}), it can be verified that
\begin{eqnarray}
\mathcal{T}^{t}_q(|\phi\rangle_{AB})=f_q(C(|\phi\rangle_{AB})),
\label{relation01}
\end{eqnarray}
where the function $f_q(x)$ is analytic and defined as
\begin{eqnarray}
f_q(x)&=&\frac{2[1-(\frac{1+\sqrt{1-x^2}}{2})^q-(\frac{1-\sqrt{1-x^2}}{2})^q]}{q-1}.
\label{}
\end{eqnarray}
This means the $\mathcal{T}^{t}_q$-entropy entanglement for qubit systems can be reduced to the Tsallis entropy entanglement. From Ref.\cite{Kim2010T}, we get a functional relation as
\begin{eqnarray}
\mathcal{T}^{t}_q(\rho_{AB})=f_q(C(\rho_{AB}))
\label{}
\end{eqnarray}
for a bipartite two-qubit mixed state $\rho_{AB}$ on Hilbert space $\mathcal{H}_A\otimes\mathcal{H}_B$. Thus, a general monogamy of the Tsallis entropy entanglement in multiqubit systems is naturally inherited by the $\mathcal{T}^{t}_q$-entropy entanglement \cite{Kim2010T,Luo2016}. However, the monogamy of multipartite higher-dimensional systems may fail for the $\mathcal{T}^{t}_q$-entropy.

\emph{Example 6.} Consider a $4 \otimes2 \otimes 2$ dimensional system in the pure state  $|\Psi\rangle_{ABC}$ given in Example 3.

\begin{figure}[htb]
\begin{center}
\resizebox{210pt}{150pt}{\includegraphics{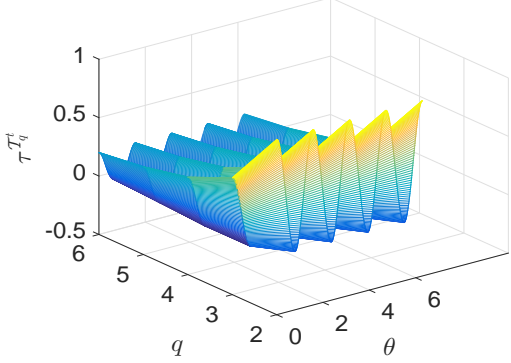}}
\end{center}
\caption{\small (Color online). The  indicator $\tau^{\mathcal{T}^{t}_q}$ vias the parameter $\theta$ for a $4 \otimes2 \otimes 2$ dimensional system in Example 6.}
\label{Tdualmonogamy}
\end{figure}

From Eq.(\ref{Tq1}), we get that
\begin{eqnarray}
\mathcal{T}^{t}_q(|\Psi\rangle_{A|BC})=\frac{2^{q+1}-c-d}{2^{q-1}(q-1)},
\end{eqnarray}
where $c=\alpha^{2q-2}+\beta^{2q-2}$, and $d=(2-\alpha^2)^{q-1}+(2-\beta^2)^{q-1}$. According to Eq.(\ref{Tq2}), we have
\begin{eqnarray}
\mathcal{T}^{t}_q(\rho_{AB})&=&2(1-\alpha^{2q}-\beta^{2q}),
\nonumber \\
\mathcal{T}^{t}_q(\rho_{AC})&=&2(1-\frac{1}{2^{q-1}}).
 \end{eqnarray}
Thus, the``residual tangle'' of the $\mathcal{T}^{t}_q$-entropy can be calculated as
\begin{eqnarray}
\tau^{\mathcal{T}^{t}_q}=\mathcal{T}^{t}_q(|\Psi\rangle_{A|BC})-\mathcal{T}^{t}_q(\rho_{AB})-\mathcal{T}^{t}_q(\rho_{AC}).
\end{eqnarray}
As its illustrated in Fig.\ref{Tdualmonogamy}, different from the $\tau^{E^{t}}$ of the $S^t$-entropy entanglement in Fig. \ref{EgEf}, the $\tau^{\mathcal{T}^{t}_q}$ entanglement may be negative or positive, where $\alpha=\cos\theta$ and $\beta=\sin\theta$. This means that the generic monogamy does not hold for high-dimensional systems in terms of the $\mathcal{T}^{t}_q$-entropy entanglement.

\end{document}